\documentclass[prd,twocolumn,eqsecnum,showpacs,amsfonts,amssymb]{revtex4}

\usepackage{graphicx}

\usepackage{bm}

\newcommand{\beq}{\begin{equation}}
\newcommand{\eeq}{\end{equation}}
\newcommand{\beqs}{\begin{eqnarray}}
\newcommand{\eeqs}{\end{eqnarray}}
\newcommand{\lsim}{\mathrel{\raisebox{-
.6ex}{$\stackrel{\textstyle<}{\sim}$}}}

\begin{document}

\title{Study of the Six-Loop Beta Function of the 
$\lambda \phi^4_4$ Theory}

\author{Robert Shrock}

\affiliation{C. N. Yang Institute for Theoretical Physics \\
Stony Brook University, Stony Brook, NY 11794 }

\begin{abstract}

We investigate whether the six-loop beta function of the $\lambda \phi^4_4$
theory exhibits evidence for an ultraviolet zero.  As part of our analysis, we
calculate and analyze Pad\'e approximants to this beta function.  Extending our
earlier results at the five-loop level, we find that in the range of $\lambda$
where the perturbative calculation of the six-loop beta function is reliable,
the theory does not exhibit robust evidence for an ultraviolet zero.

\end{abstract}


\maketitle


\section{Introduction}
\label{intro}

There has long been interest in the renormalization-group (RG) behavior of
the $\lambda \phi^4$ field theory in $d=4$ spacetime dimensions, where $\phi$
is a real scalar field.  This theory is described by the Lagrangian \cite{alt} 
\beq
{\cal L} = \frac{1}{2}(\partial_\nu \phi)(\partial^\nu \phi) 
- \frac{m^2}{2} \phi^2 - \frac{\lambda}{4!} \, \phi^4 \ . 
\label{lagrangian}
\eeq
The coupling $\lambda$ in ${\cal L}$ is
taken to be positive for the stability of the theory.  The Lagrangian
(\ref{lagrangian}) is invariant under the global discrete ${\mathbb Z}_2$
symmetry $\phi \to -\phi$.  This theory is sometimes denoted $\lambda\phi^4_4$,
with the subscript 4 indicating the spacetime dimensionality; henceforth, this 
value of $d$ will be understood implicitly.  The sign of $m^2$ will not be
important for our analysis of the ultraviolet behavior of the theory; 
for definiteness, we assume that $m^2 > 0$.

The dependence of the running coupling $\lambda(\mu)$ on the Euclidean
energy/momentum scale, $\mu$, where it is measured, is described by the beta
function of the theory \cite{rg}, $\beta_\lambda = d\lambda/dt$, where $dt=d\ln
\mu$. (The argument $\mu$ will often be suppressed in the notation.)  The
one-loop term in this beta function has a positive coefficient, so that for
small $\lambda$, $\beta_\lambda > 0$ and hence as $\mu \to 0$, the coupling
$\lambda(\mu) \to 0$, i.e., the theory is infrared-free.  This perturbative
result is in agreement with nonperturbative approaches \cite{nonpert} and is
sometimes described as the ``triviality'' property of the theory.  One then
interprets the theory as an effective one that is applicable only over a
limited range of scales $\mu$ (e.g., \cite{weinbergbook,zinnjustinbook}).  In
this theory, as $\mu$ increases from small values in the infrared (IR) to
larger values toward the ultraviolet (UV), the running coupling $\lambda(\mu)$
increases.  If one were to retain only the one-loop term in the beta function,
then this would lead to an apparent pole in $\lambda(\mu)$ at a finite value of
$\mu$, denoted $\mu_p$.  As is well known, it would not be valid to infer the
existence of a pole in $\lambda(\mu)$ at $\mu=\mu_p$, since $\lambda(\mu)$
would become too large for the perturbative calculation to be reliable before
$\mu$ reached $\mu_p$.  Nevertheless, this provides a motivation to calculate
and analyze higher-loop terms in the beta function for this theory.

An important question is whether this beta function has a UV zero, which could
thus constitute an ultraviolet fixed point (UVFP) of the renormalization group
(RG), so that as $\mu$ increases from the infrared (IR) limit $\mu=0$ to the UV
limit $\mu \to \infty$, $\lambda(\mu)$ would increase, but would approach a finite
value, $\lambda_{_{UV}}$.  In \cite{lam} we investigated this question for the
general O($N$) $\lambda |\vec \phi|^4$ theory with a real $N$-component scalar
field $\vec \phi = (\phi_1,...,\phi_N)$, using the beta function
calculated to the highest loop order available, namely five loops. Our
conclusion from that analysis was that the beta function for the O($N$) model 
$\lambda |\vec \phi|^4$ theory does not exhibit evidence for such a UVFP. 
This finding is consistent with the view of
this theory as an effective field theory, to be applied only over a restricted
range of momentum scales $\mu$.  The $\lambda\phi^4$ theory of
Eq. (\ref{lagrangian}) is the special case of the O($N$) 
$\lambda |\vec \phi|^4$ theory with $N=1$, where the continuous global 
O($N$) symmetry is reduced to a discrete ${\mathbb Z}_2$ symmetry. 

In this paper we use the recently calculated six-loop term in the 
(${\mathbb Z}_2$-invariant) $\lambda\phi^4$ theory \cite{b6} to extend
our investigation of the question of a possible UV zero in the beta function to
the six-loop level.  For perspective, one might ask whether a
(Lorentz-invariant) infrared-free quantum field theory is known whose beta
function does exhibit a UV zero.  The answer is yes, and an example is provided
by the nonlinear O($N$) $\sigma$ model in $d=2+\epsilon$ 
spacetime dimensions. In this theory, an exact solution was obtained in the 
limit $N \to \infty$ with $\lambda(\mu) N = x(\mu)$ a fixed function of $\mu$
and yielded the beta function 
\beq
\beta_x = \frac{dx}{dt} = \epsilon x \Big ( 1 - \frac{x}{x_{_{UV}}} \Big ) 
\label{betax}
\eeq
for small $\epsilon$, where $x_{_{UV}}=2\pi\epsilon$ is a UV fixed point of the
renormalization group \cite{nlsm}.  Thus, in this nonlinear O($N$) $\sigma$
model in $d=2+\epsilon$ dimensions, the coupling $x(\mu)$ flows (monotonically)
from $x=0$ at $\mu=0$ to $x=x_{_{UV}}$ as $\mu \to \infty$.  The question that
we investigate here is whether there is evidence for a similar type of behavior
in the $\lambda\phi^4$ theory in $d=4$ dimensions at the six-loop level.

This paper is organized as follows.  In Section \ref{betafunction} we discuss
some background and list the coefficients in the beta function that we will use
for our study. In Section \ref{zeros} we investigate the question of the
presence or absence of a UV zero of the beta function up to six-loop order.
Section \ref{pades} contains a further analysis of this question of a UV zero
using Pad\'e approximants.  Our conclusions are summarized in Section
\ref{conclusions}.


\section{Beta Function and Properties of Coefficients up to Five Loops}
\label{betafunction}

The beta function $\beta_\lambda=d\lambda/dt$ has the series expansion 
$\beta_\lambda = \lambda \sum_{\ell=1}^\infty b_\ell \, a ^\ell$, where 
\beq
a \equiv \frac{\lambda}{16\pi^2} \ . 
\label{a} 
\eeq
The corresponding beta function $\beta_a = da/dt$ has the series expansion
\beq
\beta_a = a \sum_{\ell=1}^\infty b_\ell \, a ^\ell \ . 
\label{beta}
\eeq
The $n$-loop ($n\ell$) beta function, denoted $\beta_{a,n\ell}$, is
given by Eq. (\ref{beta}) with the upper limit of the loop summation index
$\ell=n$ instead of $\ell=\infty$.  Thus, $\beta_{a,n\ell}$ is a polynomial in
$a$ of degree $n+1$ having an overall factor of $a^2$.  It is convenient
to extract this factor and define a reduced beta function 
\beq
\beta_{a,r} = \frac{\beta_a}{b_1 a^2} =
 1 + \frac{1}{b_1} \, \sum_{\ell=2}^\infty b_\ell a^{\ell-1} \ . 
\label{betared}
\eeq
We denote $\beta_{a,r,n\ell}$ as the $n$-loop truncation of this series. 
Thus, $\beta_{a,r,n\ell}$ is a polynomial of degree $n-1$ in $a$.  For a table
of coefficients to be presented below it will also be convenient to define the
rescaled coefficients 
\beq
\bar b_\ell \equiv \frac{b_\ell}{(4\pi)^\ell} \ . 
\label{bellbar}
\eeq

The one-loop and two-loop coefficients in the beta function, $b_1$
and $b_2$, are independent of the scheme used for regularization and
renormalization, while the coefficients at loop order three and higher,
$b_\ell$ for $\ell \ge 3$, are scheme-dependent.  The first two coefficients 
are \cite{bgz74}
\beq
b_1 = 3 
\label{b1}
\eeq
and
\beq
b_2 = -\frac{17}{3} \ . 
\label{b2}
\eeq
As noted above, since $b_1 > 0$, it follows that for small $a$, $\beta_a$ is
positive, so that as $\mu \to 0$, $a(\mu) \to 0$, i.e., the theory is IR-free.
As $\mu$ increases, $a(\mu)$ increases.  The question to be investigated here
is whether this increase in $a(\mu)$ stops, i.e., whether $a(\mu)$ approaches a
finite value $a_{_{UV}} = \lambda_{_{UV}}/(16\pi^2)$ as $\mu \to \infty$, with
$\beta(a) \to 0$ as $a \nearrow a_{_{UV}}$, or whether, instead, $\beta_a$ has
no (reliably perturbatively calculable) UV zero, so that $a(\mu)$ continues 
to grow with $\mu$
until it passes out of the region in which $\beta_a$ can be reliably
calculated perturbatively.  Here we extend our earlier five-loop analysis of
this question in \cite{lam} to the six-loop level. 

The $n$-loop coefficients $b_n$ have been calculated for the general 
O($N$) $\lambda |\vec \phi|^4$ theory up to $n=5$ loop order in the 
$\overline{\rm MS}$ scheme \cite{msbar}.  For our present
purposes, we only need the values of these $b_n$ for the theory with $N=1$.
These coefficients at the three-, four-, and five-loop level, as calculated in 
the $\overline{\rm MS}$ scheme, are \cite{vkt,bgz74,b345,kleinertbook}
\beq
b_3=\frac{145}{8} + 12\zeta_3 \ , 
\label{b3}
\eeq
\beq
b_4 = -\frac{3499}{48} - 78\zeta_3 + 18\zeta_4 - 120\zeta_5 \ , 
\label{b4}
\eeq
and
\beqs
b_5 & = & \frac{764621}{2304} + \frac{7965}{16}\zeta_3 - \frac{1189}{8}\zeta_4
+987\zeta_5 + 45\zeta_3^2 \cr\cr
& - & \frac{675}{2}\zeta_6+1323\zeta_7 \ , 
\label{b5} 
\eeqs
where 
\beq
\zeta_s = \sum_{n=1}^\infty \frac{1}{n^s}
\label{zeta}
\eeq
is the Riemann zeta function.
If $s=2r$ is even, then $\zeta_s$ can be expressed as a rational number times 
$\pi^{2r}$, namely $\zeta_{2r}=(-1)^{r+1}B_{2r}(2\pi)^{2r}/[2(2r)!]$, where 
$B_n$ are the Bernoulli numbers;
however, we leave these $\zeta_{2r}$ in their generic form here and below.  
Recently, the six-loop coefficient has been calculated 
(in the $\overline{\rm MS}$ scheme) \cite{b6} and is
\begin{widetext}
\beqs
b_6 & = & -\frac{18841427}{11520} - \frac{779603}{240}\zeta_3 
+ \frac{16989}{16}\zeta_4 -\frac{63723}{10}\zeta_5 -\frac{8678}{5}\zeta_3^2
+ \frac{6691}{2}\zeta_6 + 162\zeta_3\zeta_4 - \frac{63627}{5}\zeta_7 \cr\cr
& - & 4704\zeta_3\zeta_5 + \frac{264543}{25}\zeta_8 -
\frac{51984}{25}\zeta_{3,5}  - 768\zeta_3^3 - \frac{46112}{3}\zeta_9 \ ,  
\eeqs
\end{widetext}
%
where 
\beq
\zeta_{s_1,s_2}=\sum_{1 \le n_1 < n_2} \frac{1}{n_1^{s_1}\, n_2^{s_2}} 
\label{zetast}
\eeq
(with $n_1$ and $n_2$ positive integers) is the double zeta value \cite{mzv}.  
In Table \ref{bnbar_auv_nloop_values} we list the values of the
$\bar b_n = b_n/(2\pi)^n$ for $1 \le n \le 6$.  The values for $1 \le n \le 5$
were used in our previous work, Ref. \cite{lam}. 

\begin{table}
\caption{\footnotesize{The first and second columns of this table list the loop
    order $n$ and the numerical values of the $n$-loop coefficients 
    $\bar  b_n=b_n/(4\pi)^n$ in the beta function for $1 \le n \le
    6$. The coefficients $\bar b_n$ with $n \ge 3$ are calculated in the
    $\overline{\rm MS}$ scheme. The third column lists values of the UV zero
    $a_{_{UV,n\ell}}$ of the $n$-loop beta function, $\beta_{a,n\ell}$ for
    $n=2,...,6$ (with $b_n$ calculated in the $\overline{\rm MS}$ scheme).
    The dash notation $-$ means that $\beta_{a,n\ell}$ has no
    physical UV zero.}}
\begin{center}
\begin{tabular}{|c|c|c|} \hline\hline
$n$ & $\bar b_n$ & $a_{_{UV,n\ell}}$ 
\\ \hline
1  & 0.2387      & $-$       \\
\hline
2  & $-0.035885$  & 0.5294   \\
\hline
3  & 0.01640     & $-$       \\
\hline
4  & $-0.01089$  & 0.2333    \\
\hline
5  & 0.009090    & $-$       \\
\hline
6  & $-0.008831$ & 0.1604   \\
\hline\hline
\end{tabular}
\end{center}
\label{bnbar_auv_nloop_values}
\end{table}
%


\section{Zeros of the $n$-Loop Beta Function up to Loop Order $n=6$ }
\label{zeros}

Here we investigate a possible UV zero, denoted
$a_{_{UV,n\ell}}$, of the $n$-loop beta function, $\beta_{a,n\ell}$.  The
double zero of $\beta_{a,n\ell}$ at $a=0$ is always present (independent of 
$n$); this is an infrared zero and hence will not be of interest here. 
We denote a UV zero of the $n$-loop beta function (or equivalently, the reduced
beta function $\beta_{a,r,n\ell}$) as $a_{_{UV,n\ell}}$, if such a zero exists. 
As background for our new six-loop study, we first review
the results from our earlier five-loop analysis \cite{lam} on the question of a
possible UV zero in the beta function. The two-loop beta function 
has a UV zero at $a=a_{_{UV,2\ell}}$, where 
\beq
a_{_{UV,2\ell}}= \frac{9}{17} = 0.5294 \ , 
\label{auv_2loop}
\eeq
where here and below, floating-point values are given to the indicated
accuracy.

In order to determine whether this two-loop UV zero is a reliable perturbative
result, one must calculate higher-loop contributions to the beta function and
ascertain if this zero is reproduced in a stable manner in these higher-loop
calculations.  This program was carried out to the five-loop level in
\cite{lam} and the answer was negative.  Since $\beta_{a,n\ell}$ at loop order
$n \ge 3$ is scheme-dependent and hence so are the zeros, it is incumbent upon
one to study the effect of a scheme transformation on this answer, and this was
done in \cite{lam}, with the result that the evidence against a UV zero
in the beta function was robust under such scheme transformations.  We recall
the results obtained in the $\overline{\rm MS}$ scheme. At the three-loop
level, $\beta_{a,3\ell}$ has no IR zero; its zeros away from the origin consist
of the complex-conjugate pair $a=0.087046 \pm 0.29084i$).  At the four-loop
level, $\beta_{a,4\ell}$ has three zeros away from the origin, namely
$a=-0.056739 \pm 0.21005i$ and $a=a_{_{UV,\ell}}=0.23332$.  As indicated, one
of these is physical and may be denoted $a_{_{UV,\ell}}$, as listed in Table 
\ref{bnbar_auv_nloop_values}, but its value is more
than a factor of two smaller than the two-loop value $a_{_{UV,2\ell}}$.  At the
five-loop level, $\beta_{a,5\ell}$ does not have any physical zeros away from
the origin; instead, its four such zeros consist of the two complex-conjugate
pairs $-0.094402 \pm 0.14585i$ and $0.14208 \pm 0.12127i$.  The physical zeros
of these $n$-loop beta functions up to loop order $n=5$ were given in Table II 
of Ref. \cite{lam}.

With the recent calculation of the six-loop coefficient, $b_6$ in \cite{b6}, 
we can analyze the zeros of the resultant six-loop beta function.  This 
function is a polynomial of degree 7 in $a$ and has the numerical form 
\beqs
\beta_{a,6\ell} & = & a^2\Big ( 3 - \frac{17}{3}a + 32.5497a^2 -271.6058a^3 
\cr\cr
& + & 2848.568 a^4 -34776.131 a^5 \Big ) \ . 
\label{beta_6loop}
\eeqs
The lower-loop functions $\beta_{a,n\ell}$ with $1 \le n \le 5$ are the
corresponding truncations of this function with degree $n+1$. Aside from the
double IR zero at $a=0$, this six-loop beta function has the zero 
$a_{_{UV,6\ell}}=0.16041$ as well as
two complex-conjugate pairs $a=-0.10272 \pm 0.10558i$ and 
$a=0.063473 \pm 0.14406i$.  We list this six-loop UV zero, together with the
lower-loop results, in Table \ref{bnbar_auv_nloop_values}.

A necessary condition for a perturbative computation of the beta function 
$\beta_a$ at a given $a$ to be reliable is that, for this value of $a$, the 
fractional difference 
\beq
\bigg | \frac{\beta_{a,n+1}-\beta_{a,n}}
             {(1/2)(\beta_{a,n+1}+\beta_{a,n})} \bigg | 
\label{betachange}
\eeq
should tend to decrease as the loop order $n$ increases. 
A related necessary condition for the reliability of a
perturbative calculation of a zero of the beta function is that if one
calculates the value of $a$ that yields this zero at two successive loop
orders, then (i) if this zero is present at one order, it should also be
present at the successive order, and (ii) the magnitude of the fractional 
difference between successive loop orders, $\Delta_{n,n+1}$, where 
\beq
\Delta_{n,n+k} = \frac{|a_{_{UV,(n+k)\ell}}-a_{_{UV,n\ell}}|}
{(1/2)(a_{_{UV,(n+k)\ell}}+a_{_{UV,n\ell}})} \ , 
\label{deltannk}
\eeq
should be reasonably small and should tend to decrease as the loop
order $n$ increases.  Specifically, one would expect that
$\Delta_{n,n+1}/a_{_{UV,n\ell}}$ and 
$\Delta_{n,n+1}/a_{_{UV,(n+1)\ell}}$ should be small compared with unity and
should tend toward zero with increasing loop order $n$. 
Our analysis up to the five-loop level in \cite{lam} showed
that neither of these two requirements is met for this theory. Indeed, the
fractional differences between successive-loop orders, $\Delta_{n,n+1}$, are
not usable for $2 \le n \le 4$, since $\beta_{a,n\ell}$ has no UV zero for
$n=3$ and $n=5$. 

Here we extend this analysis to the next higher-loop order, namely $n=6$ loops.
Our six-loop results confirm and extend our previous conclusion in \cite{lam}.  Although
$\beta_{a,n\ell}$ has UV zeros at loop order $n=2$, $n=4$, and $n=6$, they are
absent at loop orders $n=3$ and $n=5$, so the first condition above is not
satisfied.  Second, even if one changes the fractional difference test to
relate not successive-loop values of $a_{_{UV,n\ell}}$ but values separated by
two loop orders, i.e., $\Delta_{n,n+2}$, these fractional differences are
substantial: 
\beq
\Delta_{2,4} = 0.776 
\label{delta_24}
\eeq
and
\beq
\Delta_{4,6} = 0.370 \ . 
\label{fracdif64}
\eeq
Furthermore, none of the quantities 
\beq
\frac{\Delta_{2,4}}{a_{_{UV,2\ell}}} = 1.467 \ , 
\label{fracdif42_over_auv_2loop}
\eeq
\beq
\frac{\Delta_{2,4}}{a_{_{UV,4\ell}}} = 3.328 \ , 
\label{fracdif42_over_auv_4loop}
\eeq
\beq
\frac{\Delta_{4,6}}{a_{_{UV,4\ell}}} = 1.587 \ , 
\label{fracdif64_over_auv_4loop}
\eeq
and 
\beq
\frac{\Delta_{4,6}}{a_{_{UV,6\ell}}} = 2.309
\label{fracdif64_over_auv_6loop}
\eeq
is small compared to unity.

\begin{figure}
  \begin{center}
    \includegraphics[height=8cm,width=6cm]{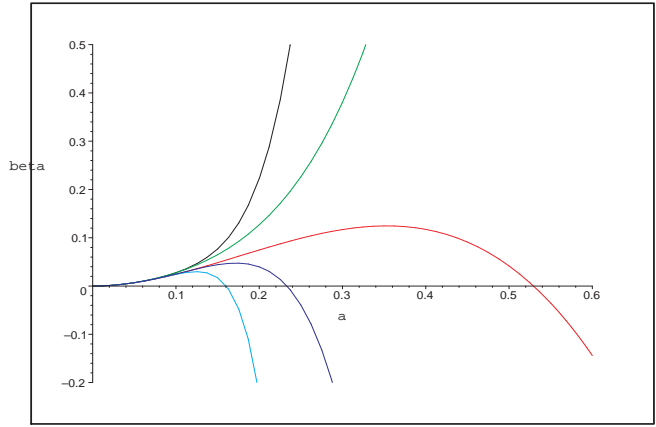}
  \end{center}
\caption{\footnotesize{ Plot of the $n$-loop $\beta$ function $\beta_{a,n\ell}$
as a function of $a$ for (i) $n=2$ (red), (ii) $n=3$ (green), (iii)
$n=4$ (blue), (iv) $n=5$ (black), and (v) $n=6$ (cyan) 
(colors in online version). 
At $a=0.16$, going from bottom to top, the curves are for 
$n=6$, $n=4$, $n=2$, $n=3$, and $n=5$.}}
\label{lam_beta_Neq1}
\end{figure}

\begin{figure}
  \begin{center}
    \includegraphics[height=8cm,width=6cm]{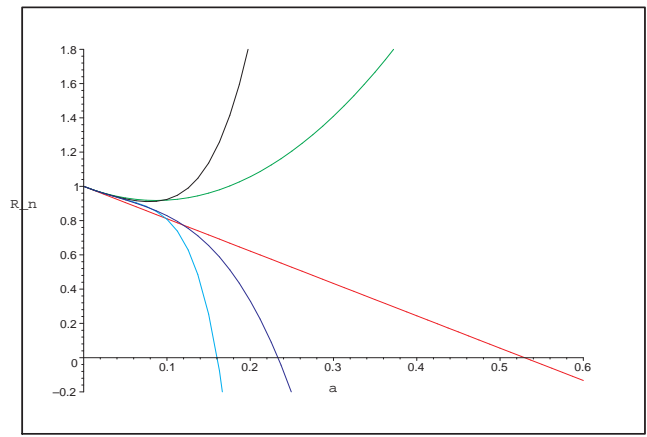}
  \end{center}                             
\caption{\footnotesize{Plot of the ratio $R_n$ of
  $\beta_{a,n\ell}$ divided by $\beta_{a,1\ell}$, as a function of $a$ for
  (i) $n=2$ (red), (ii) $n=3$ (green), (iii) $n=4$ (blue), (iv) 
  $n=5$ (black), and (v) $n=6$ (cyan) 
(colors in online version). 
At $a=0.16$, going from bottom to
    top, the curves are for $n=6$, $n=4$, $n=2$, $n=3$, and $n=5$.}}
\label{lam_betareduced_Neq1}
\end{figure}

In Fig. \ref{lam_beta_Neq1} we plot the respective $n$-loop beta functions
$\beta_{a,n\ell}$ for $2 \le n \le 6$ loops. This plot shows the
intervals in $a$ over which the calculations of $\beta_{a,n\ell}$ to the 
respective $n$-loop orders are in mutual agreement.  An alternative way to
investigate this is to plot the reduced beta function (\ref{betared}). 
We have 
\beq
\beta_{a,r,n\ell} = \frac{\beta_{a,n\ell}}{\beta_{a,1\ell}} \equiv R_n
\ . 
\label{ran}
\eeq
We plot $R_n$ in Fig. \ref{lam_betareduced_Neq1}. 


As one can see from Fig. \ref{lam_betareduced_Neq1}, the $n$-loop beta
functions $\beta_{a,n\ell}$ with $2 \le n \le 6$ only agree with each other
well over the small interval of couplings $0 \le a \lsim 0.05$; as $a$
increases beyond the upper part of this interval, they deviate from each other.
As is shown in Fig. \ref{lam_beta_Neq1}, 
the beta functions $\beta_{a,n\ell}$ with even $n=2, \ 4, \ 6$ reach maxima and
then decrease, crossing the (positive) real axis at respective values
$a_{_{UV,2}}=0.529$, $a_{_{UV,4}}=0.233$ and $a_{_{UV,6}}=0.160$ that decrease
strongly with increasing $n$, while the $\beta_{a,n\ell}$ with odd $n$ increase
monotonically as $a$ increases from zero.  The corresponding behaviors are
evident for the ratios $R_n$ shown in Fig. \ref{lam_betareduced_Neq1}.  These
results extend to the six-loop level our previous five-loop results reported in
\cite{lam} and continue the same trends observed there.  Particularly
noteworthy is our present finding that even using very high-order calculations
up to six-loop order does not significantly increase the range in $a$ in which
the beta functions calculated to adjacent-loop orders $(n,n+1)$ agree with each
other.  With this six-loop analysis, we thus confirm and strengthen our
conclusion in \cite{lam} that the zero in the two-loop beta function of the
$\lambda\phi^4$ theory occurs at too large a value of $a$ for the perturbative
calculation to be reliable.


\section{Analysis With Pad\'e Approximants}
\label{pades}

In this section we analyze the six-loop beta function for the $\lambda \phi^4$
theory using Pad\'e approximants (PAs). Since we are not interested in the
double zero in $\beta_{a,n\ell}$ at the origin, it is convenient to utilize the
reduced beta function $\beta_{a,r,n\ell}$ for this Pad\'e analysis.  The
$[p,q]$ Pad\'e approximant to $\beta_{a,r,n\ell}$ is the rational function
\beq
[p,q]_{\beta_{a,r,n\ell}} =
\frac{1+\sum_{j=1}^p \, r_j a^j}{1+\sum_{k=1}^q s_k \, a^k}
\label{pqx}
\eeq
with
\beq
p+q=n-1 \ , 
\label{pqnrel}
\eeq
where the coefficients $r_j$ and $s_j$ are independent of $a$.
At loop order $n$, we can calculate the $[p,q]_{\beta_{a,r,n\ell}}$ Pad\'e approximants with
$p+q=n-1$.  There are thus $n$ Pad\'e approximants to the $n$-loop reduced beta
function $\beta_{a,r,n\ell}$, viz., the set 
$\{ \ [n-k,k-1]_{\beta_{a,r,n\ell}} \ \}$ with $1 \le k \le n$.  Because the
value of loop order $n$ is obvious for a given Pad\'e approximant 
$[p,q]_{\beta_{a,r,n\ell}}$ from Eq.(\ref{pqnrel}), one may omit the subscript 
and write this approximant simply as $[p,q]$, and we shall do so below. 

There are several necessary requirements for a
zero of a $[p,q]$ Pad\'e approximant to be physically relevant. These include
the requirement that this zero should occur on the positive real axis in the
complex $a$ plane at a value that is not too different from $a_{_{UV,2\ell}}$
and the requirement that this zero of the PA should be closer to the origin
$a=0$ than any pole on the real positive $a$-axis, since otherwise the pole
would dominate the IR to UV flow starting at the origin. The second requirement
is clearly satisfied if a Pad\'e approximant has a denominator polynomial in
which all of the coefficients are positive.  As will be evident from the PAs to
be displayed, this positivity condition on the coefficients in the denominators
is met for all of the PAs except for the [1,3] PA at the
five-loop level and the [1,4] PA at the six-loop level,
so we only need to check the poles explicitly for these two approximants (and
neither has any relevant physical pole).

An analysis of these Pad\'e approximants up to the five-loop level in 
\cite{lam} confirmed the conclusions reached by analysis of the zeros of
$\beta_{a,r,n\ell}$ for $2 \le n \le 5$, namely evidence against a stable,
reliably calculable UV zero in the beta function.  Here we 
display the actual Pad\'e approximants up to the five-loop
level (which were not given explicitly in \cite{lam}) for reference, and,
furthermore, we analyze the Pad\'e approximants to the reduced six-loop beta function,
$\beta_{a,r,6\ell}$. For comparison with our new six-loop results, 
we recall the values of the zeros of the PAs up to the five-loop level from
\cite{lam}.  A general result that we established up to the
five-loop level in \cite{lam} was that none of the PAs has any physical pole,
i.e. a pole occurring at a real positive value.  We extend this result to the
six-loop level here.  This is clear from an inspection of the
coefficients of the denominator polynomials; these are all positive, which
immediately proves that the denominators never vanish for any real positive 
value of $a$.  

At the three-loop level, one can calculate the following set of Pad\'e
approximants to the reduced beta function $\beta_{a,r,3\ell}$:
 $\{ [2,0], \ [1,1], \ [0,2] \}$.  The $[2,0]$ PA
is $\beta_{a,r,3\ell}$ itself, and the $[0,2]$
approximant has no zeros and is therefore not useful for the analysis of a
possible UV zero. This leaves the [1,1] PA to be examined. 
In \cite{lam} we gave the zeros and poles of this PA and mentioned 
it has no physical UV zero.  Here we list it for reference: 
\beq
[1,1] = \frac{1+3.85517a}{1+5.74406a} \ . 
\label{bpade11}
\eeq
This [1,1] PA has an unphysical zero at $a=-0.2594$.

At the four-loop level, one can calculate the following set of Pad\'e
approximants to the reduced beta function $\beta_{a,r,4\ell}$:
 $\{ [3,0], \ [2,1], \ [1,2], \ [0,3] \}$.  The $[3,0]$ PA
is $\beta_{a,r,4\ell}$ itself, and the $[0,3]$
approximant has no zeros and is therefore not useful here. 
In \cite{lam} we gave the zeros and poles of these
approximants and noted that none of these was a physical UV zero.  Here
we present the actual Pad\'e approximants for reference.  They are 
\beq
[2,1] = \frac{1+6.45546a-4.91165a^2}{1+8.344345a} 
\label{bpade21}
\eeq
and
\beq
[1,2] = \frac{1+7.72950a}{1+9.61839a+7.31817a^2} \ .  
\label{bpade12}
\eeq
The [2,1] PA has an unphysical zero at $a=-0.1400$ and a UV zero at 
$a=1.4543$, which is much larger than the four-loop zero of $\beta_{a,4\ell}$
at $a_{_{UV,4\ell}}=0.2333$ and hence is not of physical relevance.  
The [1,2] PA has an unphysical zero at $a=-0.1294$. 

At the five-loop level, one can calculate the following set of Pad\'e
approximants to the reduced beta function $\beta_{a,r,5\ell}$:
 $\{ [4,0], \ [3,1], \ [2,2], \ [1,3], \ [0,4] \}$.  The $[4,0]$ PA
is $\beta_{a,r,5\ell}$ itself, and the $[0,4]$
approximant has no zeros and is therefore not useful for our analysis.
In \cite{lam} we gave the zeros and poles of these
approximants and noted that none of these was a physical UV zero.  Here
we present the actual Pad\'e approximants.  They are 
\beq
[3,1] = \frac{1+8.5989a-8.9605a^2+23.2571a^3}{1+10.4879a} \ , 
\label{bpade31}
\eeq
\beq
[2,2] = \frac{1+13.3341a+21.6066a^2}{1+15.2230a+39.51125a^2} \ , 
\label{bpade22}
\eeq
and
\beq
[1,3] = \frac{1+10.5387a}{1+12.4276a+12.6245a^2-20.4568a^3}  \ . 
\label{bpade13}
\eeq
As noted in \cite{lam}, none of these PAs has any physical zeros. The 
[3,1] PA has unphysical zeros at $a=-0.10245$ and $a=0.2439 \pm 0.6002i$;
the [2,2] PA has unphysical zeros at $a=-0.5298$ and $a=-0.08736$; and the
[1,3] PA has an unphysical zero at $a=-0.9489$. 
As a special case of the general discussion above, since the
coefficients of all terms in the denominators of the [3,1] and [2,2] PAs are
positive, it follows that neither one has any physical pole.  As was remarked in
\cite{lam}, the [1,3] PA has unphysical poles at $a=-0.46439$ and $a=-0.08986$
and a third pole at $a=1.1714$.  Since this third pole lies farther from the
origin than $a_{_{UV,n\ell}}$ with $n=2,\ 4$, one may infer that it does not
affect the RG flow from the origin in the IR to the UV and hence is not
physically relevant.

At the level of $n=6$ loops, we can calculate the following set of Pad\'e
approximants to $\beta_{a,r,6\ell}$ (a polynomial of degree 5 in $a$): 
 $\{ [5,0], \ [4,1], \ [3,2], \ [2,3], \ [1,4], \ [0,5] \}$.  The $[5,0]$ PA
is $\beta_{a,r,6\ell}$ itself, which we have already analyzed, and the $[0,5]$
approximant has no zeros and hence is not useful for our analysis.  This leaves
us with the other four PAs in the set above.
We calculate the following Pad\'e approximants to the six-loop reduced beta
function $\beta_{a,r,6\ell}$: 
\beq
[4,1] = \frac{1+10.3193a-12.2102a^2+41.9233a^3-155.757a^4} 
{1+12.2083a} \ , 
\label{bpade41}
\eeq
\beq
[3,2] = \frac{1+17.0166a+45.3789a^2-18.0872a^3}{1+18.9055a+70.2394a^2} \ , 
\label{bpade32}
\eeq
\beq
[2,3] = \frac{1+17.8537a+56.5411a^2}{1+19.7426a+82.9828a^2 +33.0754a^3} \ , 
\label{bpade23}
\eeq
and
\beqs
[1,4] & = & \frac{1+12.48863a}{1+14.3775a+16.3076a^2-34.6560a^3+ 109.7524 a^4} 
\ . \cr\cr
& & 
\label{bpade14}
\eeqs
We find that none of these Pad\'e approximants has a physical UV zero at a
value near to $a_{_{UV,6\ell}}=0.1604$. The [4,1] PA has only one physical 
UV zero, at $a=0.4675$, about 3 times larger than $a_{_{UV,6\ell}}$, as well
as unphysical zeros at $a=-0.085055$ and $a=-0.05663 \pm 0.3978i$. 
The [3,2] PA has a UV zero at the value $a=2.846$, which is too large to be 
trustworthy and, moreover, is much larger than $a_{_{UV,6\ell}}$.  
It also has two unphysical zeros at 
$a=-0.2637$ and $a=-0.07366$. The [2,3] PA has unphysical zeros at
$a=-0.2430$ and $a=-0.07279$. Finally, the [1,4] PA has a zero
at the unphysical value $a=-0.08007$.  Since the [4,1], [3,2], and [2,3] PAs
have denominators with completely positive coefficients, it is clear that they
do not have any poles on the positive real $a$ axis.  The [1,4] PA also has
only unphysical poles, which occur at $a=-0.301245$, $a=-0.07784$, 
and $a = 0.3474 \pm 0.5175i$. 

Thus, our analysis with Pad\'e approximants of the six-loop beta function
yields the same conclusion as our analysis of the beta function itself, namely
that there is no evidence for a stable, reliably perturbatively calculable UV
zero up to this six-loop level.


\section{Conclusions} 
\label{conclusions}

In this paper we have investigated whether the real scalar field theory with a
$\lambda \phi^4$ interaction (in four spacetime dimensions) exhibits evidence
of an ultraviolet zero in the beta function. Using the recently calculated
six-loop term $b_6$ from \cite{b6}, our present study extends our previous
five-loop study in \cite{lam} to the six-loop level.  From a study of the
six-loop beta function itself together with an analysis of Pad\'e approximants,
we conclude that this beta function does not exhibit evidence for a stable,
reliably perturbatively calculable UV zero to the highest loop order, namely
six loops, to which it has been computed. 


\begin{acknowledgments}

This research was partly supported by the U.S. National Science Foundation 
Grant NSF-PHY-16-1620628. 

\end{acknowledgments}



\begin{thebibliography}{99}
\bibitem{alt}
%
Another common normalization uses ${\cal L}_{int} =
-(1/4)\lambda_{alt} \, \phi^4$. These are equivalent, with
$\lambda = 6\lambda_{alt.}$.
\end{thebibliography}
\end{document}